\documentclass[%
 aps,
 pra,
 superscriptaddress,
 sd,%
 amsmath,amssymb,
 reprint,%
showpacs
]{revtex4-1}

\usepackage{color}
\usepackage{graphicx}
\usepackage{dcolumn}
\usepackage{bm}

\begin{document}


\title{How light absorption modifies the radiative force on a microparticle in optical tweezers}

\author{W. H. Campos}
\email{Corresponding author: warlley.campos@ufv.br}
\author{J. M. Fonseca}
\author{J. B. S. Mendes}
\author{M. S. Rocha}
\author{W. A. Moura-Melo}
\affiliation{Departamento de
F\'isica, Universidade Federal de Vi\c{c}osa, 36570-900, Vi\c{c}osa, Minas Gerais, Brazil.}%


\date{\today}

\begin{abstract}

Reflection and refraction of light can be used to trap small dielectric particles in the geometrical optics regime. Absorption of light is usually neglected in theoretical calculations, but it is known that it occurs in the optical trapping of metallic particles. Also, recent experiments with semi-transparent microparticles have shown that absorption of light is important to explain their optically induced oscillations. Here, we propose a generalization of Ashkin's model for the radiative force exerted on a spherical bead, including the contribution due to attenuation/absorption of light in the bulk of the particle. We discuss in detail the balance between refraction, reflection and absorption for different optical parameters and particle sizes. A detailed example is provided in order to clarify how the model can be applied, and it is obtained that the radiative force can either increase or decrease with absorption, depending on the particle size. Our findings contribute to the understanding of optical trapping of light-absorbing particles, and may be used to predict whenever absorption is important in real experiments.
\end{abstract}

\maketitle

\section{Introduction}

The remote trapping and manipulation of micro-sized objects using light was first proposed and accomplished by A. Ashkin and collaborators \cite{Ashkin1,Ashkin2,Ashkin}, whom demonstrated an application for their findings by manipulating viruses and bacteria in aqueous solution \cite{Ashkin3}. Such investigations gave rise to the rapidly growing field of \emph{optical tweezers} (or optical trapping). Nowadays, optical tweezers techniques are established as powerful tools in different areas of physics, biology and chemistry, contributing mainly to the study of biological cells and macromolecular systems, such as single biopolymers and cell membranes \cite{Bustamante,Ashkin4,Svoboda,Rocha1,Gao,Hayashi,Jiang}. Optical ``tractor beams'' have also been used for microscopic transport of airborne  particles \cite{Shvedov1}, contributing to studies in aerosol science \cite{Zhang,Pan,Wang,Gong,Shvedov2,Shvedov3,Lin,Redding}. The optical torque exerted on small chiral particles for generic optical fields was calculated recently \cite{Chen}, and the optical tweezers technique was used to probe Casimir interactions at micro and nanometer scale \cite{Ether,Xu}. Broad review articles and a step-by-step guide to realization of optical tweezers can be found in Refs. \cite{Molloy,Grier,Neuman,Dholakia,Pesce}.

In order to trap small particles with optical tweezers, one has to consider the optical properties of the material that they are made, since these properties are critical to determine the resulting forces acting on the object. When subjected to a highly focused Gaussian light beam, dielectric particles are known to yield stable trapping in three dimensions, which was theoretically described by R. Gauthier, S. Wallace and A. Ashkin in geometrical optics regime \cite{Gauthier,Ashkin} ($\lambda<<l$, where $\lambda$ is the laser wavelength and $l$ is the particle radius), by J. Gordon in Rayleigh regime \cite{Gordon} ($\lambda>>l$), and by P. Maia Neto, H. Nussenzveig and coworkers in Mie regime \cite{Mazolli,Dutra} ($\lambda\sim l$). Metallic particles can be trapped under very special conditions. In Rayleigh regime, the restoring force is proportional to the particle electric polarizability, so that very small metallic particles can be trapped in three dimensions with greater efficiency than dielectric ones \cite{Svoboda2}. In Mie and geometrical optics regimes, it is known that metallic particles need to be trapped against a substrate, i.e. the optical tweezing stability is only bidimensional \cite{Sato,Furukawa,Ke}.  Alternatively, metallic Mie particles can be trapped by subjecting them to an inverted ``doughnut'' laser beam \cite{Lewittes}. A theoretical description for the trapping of metallic particles is given in Ref. \cite{Ke}, where the authors have considered an spherical particle that reflects 100$\%$ of the incoming light.

Regarding optical tweezers, the absorption of light by particles in geometrical optics regime is usually neglected in theoretical models and avoided in experiments, since it results in generation of the so-called radiometric forces \cite{Ke,Lewittes}. These are originated from temperature gradients in the medium where the beads are immersed in, and they are called photophoretic forces when caused by light \cite{Ashkin1,Pluchino}. However, radiometric effects have become not only important, but also useful in optical control and manipulation of microparticles. For example, inhomogeneous heating of the medium was recently reported as an efficient method for optothermal trapping of particles in a hollow-core photonic crystal fiber \cite{Schmidt,Dholakia1}. In usual Gaussian beam setups, the photophoretic forces due to absorption of light by the particles tend to scatter them from the optical region, but they are exactly the same forces responsible to levitate metallic particles in a ``doughnut'' laser beam \cite{Lewittes}. Although for metallic particles the radiometric forces highly overcomes the radiative (radiation pressure $+$ gradient) forces, if the particle is semi-transparent they can compete, eventually yielding to quasi-periodic oscillations, as recently observed for topological insulator (Bi$_2$Te$_3$ and Bi$_2$Se$_3$) particles in aqueous solution \cite{Campos}.

Optical trapping and manipulation of light-absorbing particles have been studied experimentally in different configurations \cite{Wang,Redding,Shvedov1,Shvedov2,Zhang}. It has been also studied analytically in Rayleigh regime \cite{Svoboda2}, and by finite-difference time-domain simulations for particles in Lorentz-Mie regime \cite{Jia}. However, to our best knowledge, an analytical expression for the radiative force exerted on an absorbing particle in the geometrical optics regime is still lacking. Whenever light is absorbed by a particle subjected to a laser beam, a ``microscopic'' description of the exerted forces requires a clear distinction between thermally originated and radiative forces. In this work, we propose a generalization of the usual model accounting for the radiative force exerted on a spherical particle by a single light ray in the geometrical optics regime \cite{Gauthier,Ashkin}. The absorption/attenuation of light is included in such a way that the model captures the limiting cases of fully transparent (dielectric) and fully reflective (metallic) particles \cite{Gauthier,Ashkin,Ke}, with the novelty of accounting for any intermediary value of the absorption coefficient, namely the case of semi-transparent particles made from semiconductor materials. As a direct example, we calculate the total radiative force exerted by a Gaussian laser beam on a particle highly out of the laser focus, a setup similar to that recently reported in the work of Ref. \cite{Campos}, and discuss upon the modifications on this force caused by different values of the absorption coefficient. We also discuss how the particle weight becomes comparable to the optical forces if its density is higher enough, which is the case of the topological insulator microparticles used in Ref. \cite{Campos}, and a number of semiconductor materials where light absorption is expected to be important.

\section{The model: Spherical absorbing particle}

\begin{figure}[h]
\includegraphics[width=\linewidth]{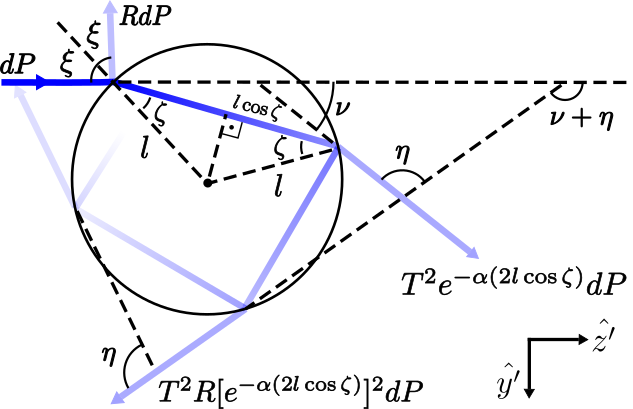}
\caption{(Color online) Light ray interacting with an absorbing particle. Besides the initial reflection and refraction suffered when it reaches the particle, the ray splits again every time it reaches the inside surface. Furthermore, it suffers an attenuation when propagating throughout the bulk of the particle to the opposite side. $\xi$ and $\zeta$ are the incidence and refraction angles, respectively, and $l$ is the sphere radius. See text for discussion.}
\label{figure1}
\end{figure}

As usual in geometrical optics regime, we follow the main ideas of Ref. \cite{Ashkin} to obtain the radiative force exerted by a light ray on a spherical bead.  In their model, the authors obtain the radiative force by computing the variation of the linear momentum of light, as a result of the interaction with the particle surface. They do not consider absorption of light by the particle, so that only reflection or refraction at the surface can modify the linear momentum of the light ray. 

Here, we consider a single ray of power $dP$ interacting with an absorbing spherical bead (Fig. \ref{figure1}). Every time the ray reaches the bead surface, a fraction $R$ of the incoming power is reflected and a fraction $T=1-R$ is transmitted through the interface (i.e. $R$ is the reflectivity and $T$ is the transmissivity at the surface). As a result of the reflection and refraction, linear momentum will be transfered from the light ray to the bead, exerting a force whose magnitude and direction depends on the angle of incidence, $\xi$. If absorption is not considered, $100\%$ of the power transmitted through the surface reaches the opposite side. Say, the particle is supposed to be perfectly transparent, which works well for dielectric beads generally made of polystyrene. However, for general materials the absorption of light cannot be neglected, and we expect that only a fraction $e^{-\alpha (2l\cos\zeta)}$ of the transmitted power will travel across the absorbing particle (Beer's law), where $l$ is the particle radius and $\zeta$ is the angle of refraction. $\alpha$ is the absorption coefficient of the material, which can be related to the extinction coefficient, $\kappa$, by $\alpha=4\pi\kappa/\lambda$ \cite{Fox}.

Once the incident light ray penetrates the interface, there will be new reflections/transmissions every time it interacts with the inside surface (Fig. \ref{figure1}). In each interaction, a fraction $T^2R^n\left[\exp^{-\alpha(2l\cos\zeta)}\right]^{n+1}$ of the initial power leaves the particle in a different direction than that of the incident ray ($n$ accounts for the number of times the ray was reflected before leaving the particle), i.e., there is also a transfer of linear momentum from the ray to the particle due to multiple reflections/transmissions.

We choose the $z'$-axis to be in the direction of the incident ray, and the $y'$-axis to be in the plane of incidence, which contains both the light ray and the center of the particle. The total force on the sphere is obtained by summing up the individual contributions of each ray-surface interaction, with $d\vec{F}_{particle}=-d\vec{F}_{ray}$. Details on the calculations are provided in Appendix A, but the final result can be written as \cite{Ashkin,Rocha}:
\begin{equation}\label{dF}
d\vec{F}=\frac{n_m}{c}[\mathrm{Re}(Q_t)\hat{z'}+\mathrm{Im}(Q_t)\hat{y'}]dP,
\end{equation}
where $n_m$ is the refractive index of the surrounding medium, $c$ is the speed of light, and 
\begin{equation}\label{eq:Q}
Q_t=1+R\exp(2i\xi)-T^2\frac{\exp[2i(\xi-\zeta)]}{e^{2l\alpha\cos\zeta}+Re^{-2i\zeta}}.
\end{equation}

This result covers the two limiting cases studied previously in the literature. Indeed, taking $\alpha=0$, one recovers the usual expression of $Q_t$ for fully transparent/dielectric beads \cite{Rocha}. In turn, taking $\alpha=+\infty$ one recovers the expression for metallic particles reflecting all the incoming light, say, $Q_t=1+R\exp(2i\xi)$ \cite{Ke}.

To get intuition about when consideration of the absorption coefficient is important, let us investigate the skin depth, which can defined as $\delta\equiv1/\alpha$. The skin depth gives an estimative of the distance traveled by light before its power falls to $1/e$ of its value at the surface. Particles used in optical tweezers are usually a few micrometers large, and transparent materials have $\delta=+\infty$, so that light can travel a distance much larger than the particle size before being considerable attenuated. In such case the transfer of momentum due to absorption is minimal and can be safely neglected, as usual. Otherwise, if the particle is metallic, $\delta=0$ (or usually a few nanometers). This is equivalent of considering a non-absorptive particle with fully reflective surface, i. e. light cannot reach the bulk, and therefore, cannot be attenuated. Here, transfer of momentum due to both refraction and absorption is minimal and can be neglected.  A more interesting scenario arises when we consider a skin depth comparable to the particle radius, $\delta\sim l$. Semiconductors materials, for example, can have a skin depth in the order of micrometers, making the microparticles semi-transparent. In such case, absorption of light by the particle is significant and should not be neglected in theoretical calculations. In this case, a small change in the absorption coefficient can lead to an appreciable modification in the radiative force. This will be discussed in more details in Section \ref{results}.

\begin{figure}[h]
\includegraphics[width=\linewidth]{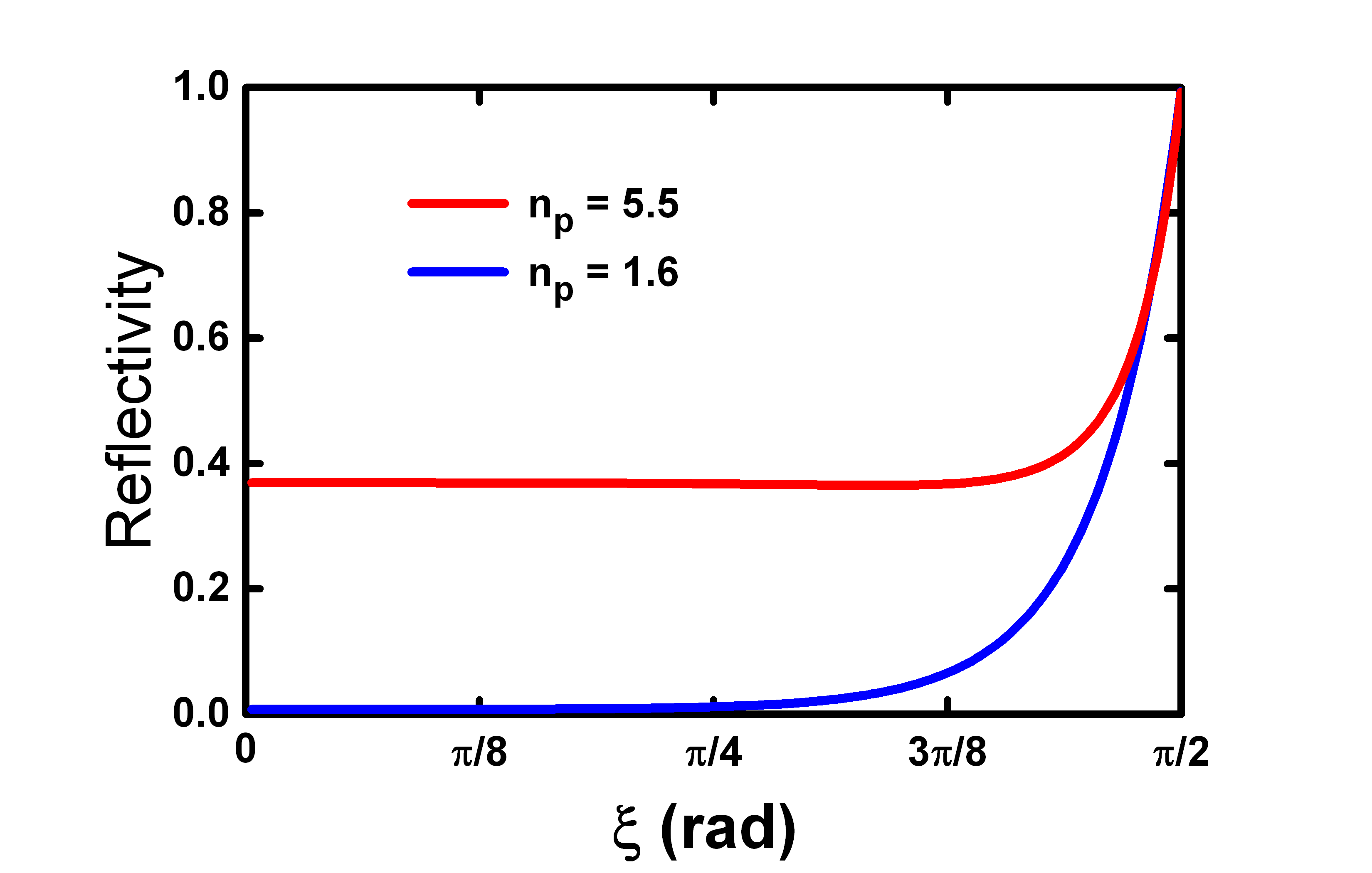}
\caption{(Color online) Surface reflectivity, $R$, as function of incident angle, $\xi$. Higher refractive index of the particle, $n_p$, leads to higher reflectivity. Also, light is more reflected for tangent angles of incidence, $\xi\rightarrow\pi/2$. We take $n_m=1.3$, the refractive index of deionized water.}
\label{img:Reflectivity}
\end{figure}

Now, consider a linearly polarized light ray and take the average over the two laser polarizations, TE and TM. Assuming no preferential polarization, the reflectivity $R$ can be written as \cite{Rocha,Jackson}:
\begin{equation}
R(\xi,\zeta)=\frac{1}{2}\left[\dfrac{\sin(\xi-\zeta)}{\sin(\xi+\zeta)}\right]^2+\frac{1}{2}\left[\dfrac{\tan(\xi-\zeta)}{\tan(\xi+\zeta)}\right]^2.
\end{equation}%
The refraction angle reads $\zeta(\xi)=\arcsin\left[\frac{n_m}{n_p}\sin\xi\right]$ (Snell's law), where $n_p$ is the refractive index of the particle. Fig. \ref{img:Reflectivity} shows the behavior of the reflectivity $R(\xi,\zeta)=R(\xi)$ for $0\leq\xi<\pi/2$ and different values of $n_p>n_m$. For polystyrene particles (usually adopted in experiments), which have $n_p\approx1.6$, light is mostly transmitted throughout the material. The exception is for $\xi\rightarrow\pi/2$ (tangent angles), where light is mostly reflected by the surface. However, in particles with higher index of refraction, such as Bi$_2$Te$_3$ and Bi$_2$Se$_3$ topological insulators (typical value of $n_p\approx5.5$ \cite{Yue}), a considerable amount of light is reflected even for small angles, generating radiation pressure. If the material has a very high refractive index, reflectivity is close to one. Therefore, the surface does not allow penetration of light rays to the bulk, and the particles should not distinguish between different values of the bulk absorption coefficient $\alpha$.
	
\section{Results and Discussion}\label{results}
	
\begin{figure*}
\includegraphics[width=.95\linewidth]{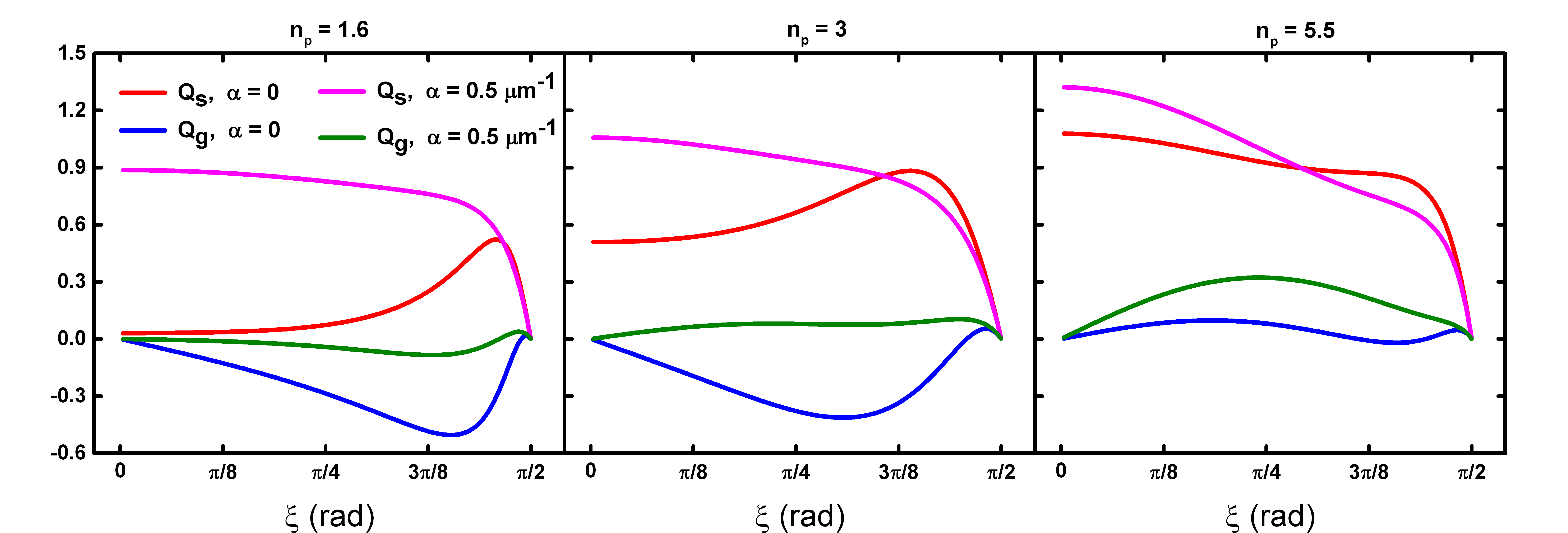}
\caption{(Color online) Plots of $Q_s$ and $Q_g$ as functions of the incidence angle, $\xi$, for different values of the refractive index, $n_p$, and absorption coefficient, $\alpha$. As $n_p$ increases, the influence of absorption in the bulk is minimized, since a significant fraction of the light is reflected by the surface and cannot be attenuated in the bulk. We have used $n_m=1.3$, and $l=2.1\,\mu$m for the sphere radius.}
\label{img:balance}
\end{figure*}

The balance between refraction, reflection and absorption can be better understood by studying the real and imaginary components of the adimensional factor $Q_t$. As usual in the literature, we define $Q_s\equiv\textrm{Re}(Q_t)$ and $Q_g\equiv\textrm{Im}(Q_t)$. The former corresponds to the component of the force along $z'$-axis, and indicates the strength in which the ray pushes the particle along its direction of incidence. The latter corresponds to the component of the force along $y'$-axis and indicates the strength in which the bead is attracted ($Q_g<0$) or deflected ($Q_g>0$) along the transverse direction. Fig. \ref{img:balance} shows the plots of $Q_s$ and $Q_g$ as function of the incident angle $\xi$ for a particle with radius $l=2.1\,\mu$m and different values of $n_p$ and $\alpha$. 

The low reflectivity of particles with small refractive index allows a great amount of light to be transmitted through the surface. As a consequence, these particles are very sensitive to different values of the absorption coefficient. In fact, for $n_p=1.6$, Fig. \ref{img:balance} (left panel) shows that there are significant changes in the behavior of $Q_s$ and $Q_g$ for different values of $\alpha$. Note the increasing in the adimensional force along $z'$-axis, $Q_s$, as the absorption coefficient changes from $\alpha=0$ to $\alpha=0.5\,\mu$m$^{-1}$, which is expected since absorption must generate radiation pressure. If $\alpha=0$, $Q_g<0$ for all values of $\xi$, and the force along $y'$-axis is attractive, as known for polystyrene particles. However, for $\alpha=0.5\,\mu$m$^{-1}$ such attraction is weakened by the absorbing bulk of the particle. In this case, the skin depth $\delta=\alpha^{-1}=2\,\mu$m is smaller than the particle diameter ($4.2\,\mu$m), so the light ray is attenuated before all multiple reflections can contribute significantly to $Q_g$. 

	For a particle with intermediary refractive index, say $n_p=3$ (Fig. \ref{img:balance}, center panel), $Q_s$ is higher than in the previous example even for $\alpha=0$. This is a consequence of the higher reflectivity, $R$, which allows the light to be only partially transmitted through the surface. Nevertheless, $\alpha=0.5\,\mu$m$^{-1}$ implies in an increasing of the adimensional force along $z'$ -axis, $Q_s$, since a fraction of the light penetrates the bulk in order to be attenuated. Notice that in this case, $Q_g<0$ if $\alpha=0$, but an absorption coefficient of $\alpha=0.5\,\mu$m$^{-1}$ eliminates the attractive behavior along $y'$-axis. 
	
	For higher values of $n_p$, a greater amount of light is reflected by the surface, and absorption does not change the behavior of $Q_t$ in a significant way, since only a small fraction of light penetrates into the bulk (Fig. \ref{img:balance}, right panel). Note that in this case ($n_p=5.5$), even if $\alpha=0$ we have $Q_g>0$ for most values of $\xi$, i.e. reflection assures that the particle will not be attracted along $y'$-axis . It can be shown that in the limit of very high refractive index (say $n\gtrsim20$, in our calculations) there are virtually no distinctions between $Q_t$ for $\alpha=0$ and for $\alpha\neq 0$, since the interface reflects all the incoming light. 
	
	It is evident, from the above results, that when considering interaction of a light ray with a spherical bead, the main consequences of a non-zero value of the absorption coefficient is to partially suppress the multiple reflections/transmissions taking place in the inside surface. Such a suppression can become total if the skin depth is much lower than the particle radius, $\delta<<l$.

\subsection*{Radiative force exerted by a focused Gaussian beam}

\begin{figure}[h]
\includegraphics[width=.8\linewidth]{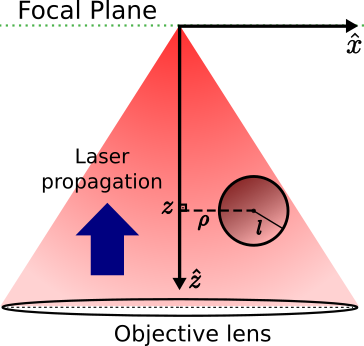}
\caption{(Color online) Illustration of a focused Gaussian light beam and definition of the coordinate system. The spherical particle is placed a distance $z$ below the focal plane, and a distance $\rho$ from the optical axis. The laser propagates from the objective lens to the origin of the coordinate system.}
\label{figure3}
\end{figure}

As a direct example of how the model can be applied, let us consider a focused Gaussian laser beam propagating along z-axis, toward the focal plane (Fig. \ref{figure3}). This is a very common setup, found in many applications of optical tweezers \cite{Rocha,Nussenzveig,Spesyvtseva}. According to the experimental results reported in Ref. \cite{Campos}, if an absorbing particle is placed close enough to the focal plane, the radiometric force is very high and completely overcome the radiative. Therefore, we are interested in a situation where the bead is placed highly out of the beam focus. Fig. \ref{figure3} depicts the spherical particle at a distance $z$ below the focal plane, while the origin of the coordinate system was chosen to be at the beam focus. We set the $z$-axis along the optical axis, toward the objective lens. The $x$-axis was chosen so that the the center of the sphere lies in the $x$-$z$ plane, and $\rho$ is the radial distance from the optical axis to the geometrical center of the sphere.

The procedure used to calculate the total radiative force exerted on the spherical bead is described in Appendix B. In summary, the force exerted by a single ray is calculated using Eq. (\ref{dF}), and integration over all the rays reaching the sphere gives the final result. In order to perform the numerical calculations, we set $P_t=15.4\,$ mW for the total laser power, and $\sigma=0.136\,$mm for the beam waist before it passes through the objective. The focal length and the refractive index of the medium are taken $f=0.375\,$mm and $n_m=1.3$, respectively. These are typical values from experimets with optical tweezers.

\begin{figure}[h]
\includegraphics[width=.9\linewidth]{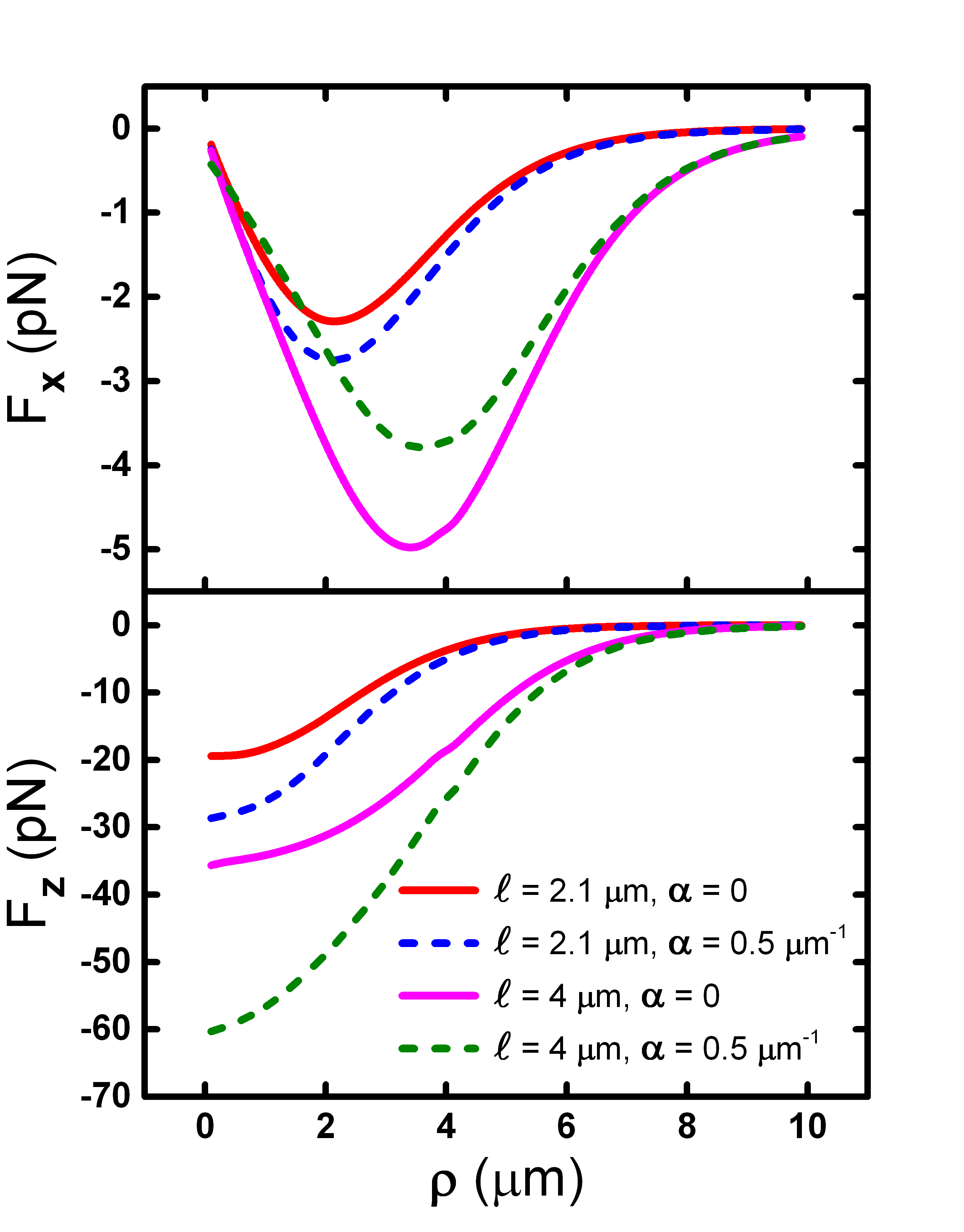}
\caption{(Color online) Radial and azimuthal components of the optical force along $x$-axis and $z$-axis, respectively, as function of the radial distance $\rho$. We set $n_p=3$ for the refractive index of the bead. The top panel shows the behavior of the radial force, $F_x$, for different values of the particle radius, $l$, and absorption coefficient, $\alpha$. The bottom panel shows the behavior of the azimuthal force, $F_z$, for the same parameters. See text for discussion.}
\label{img:opticalforces}
\end{figure}

Figure \ref{img:opticalforces} (top panel) shows the radial component of the radiative force along $x$-axis for a particle of refractive index $n_p=3$, and different values of the absorption coefficient, $\alpha$, and radius, $l$. Observe that for smaller particles ($l=2.1\,\mu$m), a higher value of $\alpha$ implies in an increasing of the radial force pushing the particle toward the optical axis, similar to the restoring force driven by reflection of light in the surface of metallic particles, when trapped in two dimensions \cite{Sato}. This can be understood by the fact that all the incoming rays are directed toward the focus of the light beam, and their linear momentum is transfered to the particle through absorption/reflection. Surprisingly, the situation is inverted for bigger particles ($l=4\,\mu$m). In this case, a higher value for $\alpha$ decreases the magnitude of $F_x$, weakening the attraction toward the optical axis. This result suggests that attraction of bigger particles is driven mainly by the multiple reflections/transmissions at the inside surface, and suppression of such interactions by absorption of light in the bulk causes the magnitude of the radial force to decrease. This result is consistent with the known fact that tridimensional stable trapping of very small metallic particles (Rayleigh regime) are more efficient than that of dielectric ones \cite{Svoboda2}, while bigger metallic particles cannot be stably trapped in three dimensions \cite{Sato}.

Figure \ref{img:opticalforces} (bottom panel) shows the azimuthal component of the optical force along $z$-axis for different values of the absorption coefficient, $\alpha$, and the particle radius, $l$. No inversion is observed for bigger particles, i.e. for both $l=2.1\,\mu$m and $l=4\,\mu$m, a higher value of the absorption coefficient $\alpha$ implies in an increasing at the magnitude of the force pushing the particle toward the focal plane (observe in Fig. \ref{figure3}-a) that the $z$-axis points down, so a negative value of $F_z$ means a force pointing up to the focal plane). 

\begin{figure}[h]
\includegraphics[width=.9\linewidth]{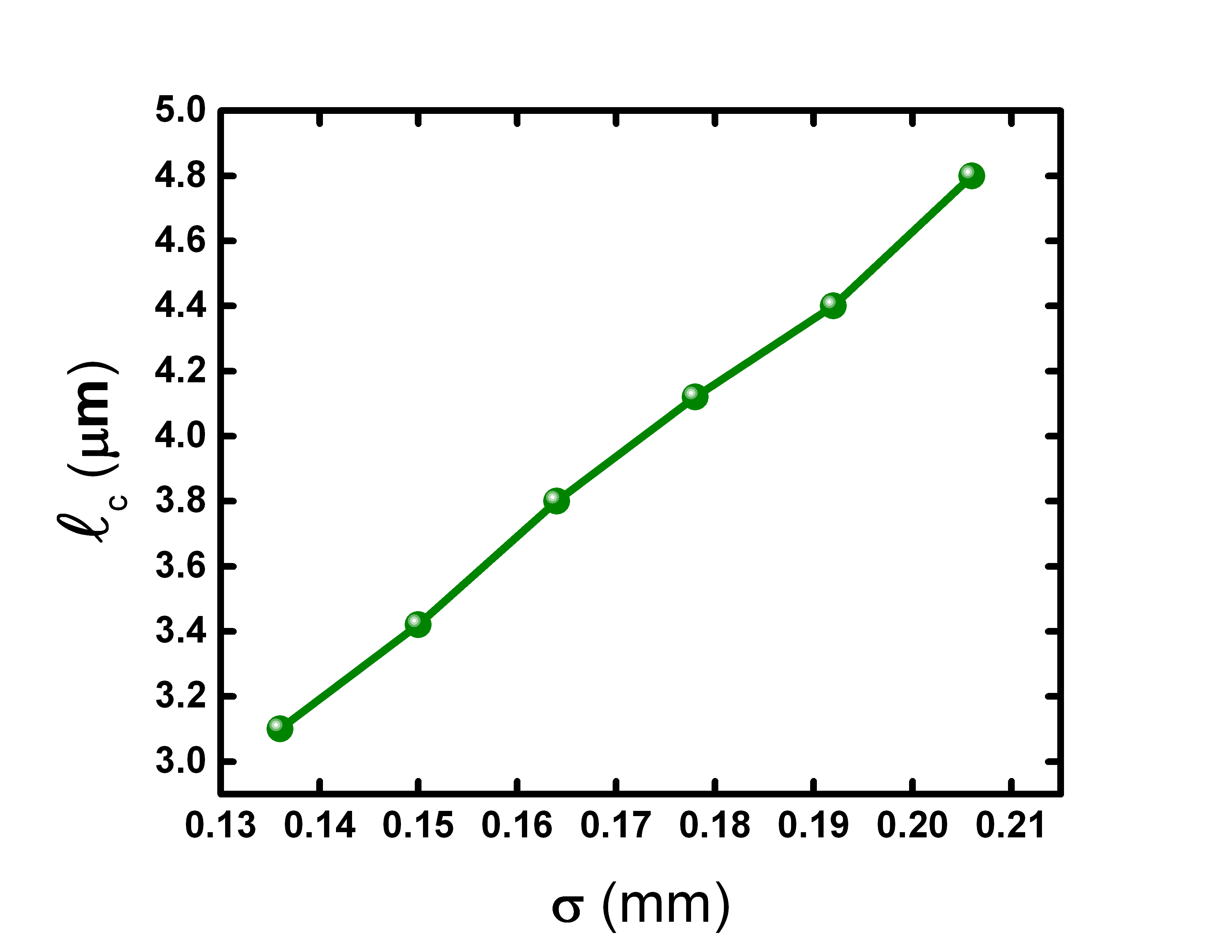}
\caption{(Color online) Critical radius, $l_c$, as function of the laser beam waist, $\sigma$. Notice that the increasing is approximately linear.}
\label{raiocritico}
\end{figure}

From Fig. \ref{img:opticalforces} (top panel), it is evident that there must be a critical radius $l_c$ where the behavior of the force with absorption coefficient is reversed. Actually, in geometrical optics regime, the waist of the laser beam is expected to significantly influence the critical radius. In fact, Fig. \ref{raiocritico} shows that $l_c$ increases linearly with the beam waist, $\sigma$. At the critical radius, the absorption coefficient $\alpha$ almost does not contribute the radiative force exerted on the particle.

\begin{figure}[h]
\includegraphics[width=\linewidth]{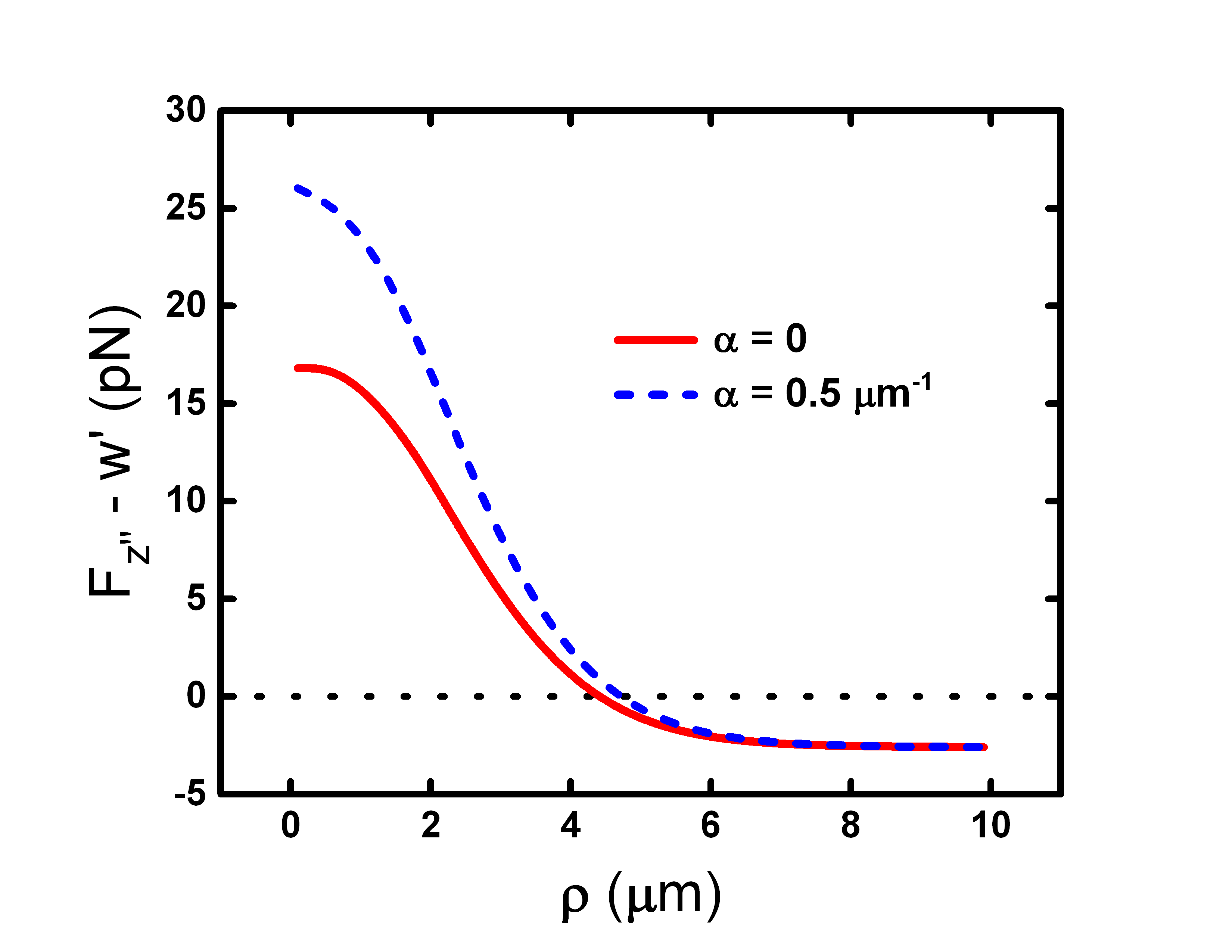}
\caption{(Color online) Resultant force ($F_{z''}-w'$) along $z''$-axis as function of the radial distance, $\rho$, for a particle with radius $l=2.1\,\mu$m, $n_p=5.5$, and density $\sim 7.85\times 10^3\,$kg/m$^3$ (as known for Bi$_2$Te$_3$ TI's). The particle is pushed toward the focal plane for positive values of the resultant force, while for negative values the weight overcomes the optical force. The same behavior is observed for both $\alpha=0$ and $\alpha=0.5\,\mu$m$^{-1}$.}
\label{plotpeso}
\end{figure}

It is important to say that we are not describing any particle dynamics here, since this demands inclusion of radiometric forces. Nevertheless, in what follows we discuss about how the buoyancy and weight of denser particles can influence the optically induced oscillations reported in Ref. \cite{Campos}. Consider a particle of radius $l\sim2.1\,\mu$m, which leads to a buoyancy around $\sim 0.38\,$pN. If the particle is made from polystyrene (density $\sim$ $1.04\times 10^3$ kg/m$^3$), it has a weight around $\sim0.4\,$pN. These forces balance each other and become negligible in most experiments, when compared to the magnitude of the other forces involved. However, here we are interested in materials with non-negligible absorption, such as Ge, Bi$_2$Te$_3$, and Bi$_2$Se$_3$.  These materials are usually denser, and the weight can overcomes the buoyancy. For instance, the density of the Bi$_2$Te$_3$ topological insulator used in Ref. \cite{Campos} reads $\sim 7.85\times 10^3$ kg/m$^3$, corresponding to a weight around $w\approx2.98\,$pN. In order to be clearer, let us define the $z''$-axis to be antiparallel to the $z$-axis, so that the weight points in its negative direction. Fig \ref{plotpeso} shows the plot of $F_{z''}-w'$ ($F_{z''}$ is the radiative force along $z''$, and $w'$ is the apparent weight) as function of the radial distance $\rho$. Near the optical axis, $F_{z''}$ overcomes the apparent weight, pushing the particle toward the focal plane. Radiometric forces would also contribute to the force along this direction. However, at larger values of $\rho$ it is the apparent weight that overcomes the optical force, and the particle must fall. This point is important in dynamic systems such as that reported in Ref. \cite{Campos}, where the particle approaches and moves away from the optical axis in a quasi-periodic motion. Such a remark leads us to predict that oscillations also take place along the vertical direction ($z$-axis), where the apparent weight acts as the restoring force and the radiative plus radiometric forces scatter the particle. This is not discussed in the work of Ref. \cite{Campos}, but certainly worth investigation in the future.

\section*{Conclusions}

A generalization of Ashkin's model \cite{Ashkin} is proposed in order to include absorption/attenuation of light into account, when considering a spherical bead under the incidence of a single light ray. Our results accounts for an arbitrary value of the absorption coefficient, including the two limiting cases previously studied in the literature, namely, perfectly dielectric \cite{Rocha} and perfectly metallic particles \cite{Ke}. Our findings suggest that attenuation effects becomes important mainly when the skin depth is comparable to the particle size. The radiative force along the light ray direction increases with the absorption coefficient, and the radiative force along the transverse direction (which is usually attractive) can be suppressed and become repulsive with increasing in the absorption coefficient. We have also calculated the radiative force exerted on a spherical particle under a Gaussian laser beam optical tweezers. Interestingly, our results suggest that the effective contribution of absorption/attenuation of light by the particle bulk highly depends on its size. If the particle is small, increasing in the absorption coefficient leads to an increasing in the radial force pushing the particle toward the optical axis. However, for bigger particles the attraction is driven mainly by the multiple reflections/transmissions at the inside surface, which are suppressed by absorption of light in the bulk, causing the magnitude of the radial force to decrease. The critical radius where this inversion occurs is shown to be proportional to the laser beam waist.


\section{Acknowledgements}

The authors thank the Brazilian agencies CAPES, FAPEMIG and CNPq for financial support.

\section*{Appendix A}

From Fig. \ref{figure1}, it can be seen that the component of the radiative force along $z'$-axis is given by \citep{Ashkin}:
\begin{equation}
\begin{split}
dF_{z'}=&\frac{n_m}{c}\Big\lbrace 1+R\cos{(2\xi)}\\-&T^2\sum_{n=0}^{+\infty}R^n\big[e^{-2l\alpha\cos{\zeta}}\big]^{n+1}\cos{(\nu+n\eta)}\Big\rbrace dP,
\end{split}
\end{equation}
where $n_m$ is the index of refraction of the medium, $n=0,1,2,3\,...$ is an integer number, and the angles $\xi, \zeta, \nu, \eta$ are indicated in Fig. \ref{figure1}. This can be easily shown to be: 

\begin{equation}
dF_{z'}=\frac{n_m}{c}\textrm{Re}\Big\lbrace 1+Re^{2i\xi}-T^2\frac{e^{i\nu}}{e^{2l\alpha\cos\zeta}-Re^{i\eta}}\Big\rbrace dP.
\end{equation}

Using the geometric relations $\nu=2\xi-2\zeta$ and $\eta=\pi-2\zeta$, one gets:

\begin{equation}\label{eq:dfz}
dF_{z'}=\frac{n_m}{c}\textrm{Re}\Big\lbrace 1+Re^{2i\xi}-T^2\frac{e^{2i(\xi-\zeta)}}{e^{2l\alpha\cos\zeta}+Re^{-2i\zeta}}\Big\rbrace dP.
\end{equation}

In turn, along $y'$-axis, we have that:

\begin{equation}
\begin{split}
dF_{y'}=&\frac{n_m}{c}\Big\lbrace R\sin (2\xi)\\-&T^2\sum_{n=0}^{+\infty}R^n\big[e^{-2l\alpha\cos{\zeta}}\big]^{n+1}\sin{(\nu+n\eta)}\Big\rbrace dP.
\end{split}
\end{equation}

Proceeding in an analogous way, one gets:

\begin{equation}\label{eq:dfy}
dF_{y'}=\frac{n_m}{c}\textrm{Im}\Big\lbrace Re^{2i\xi}-T^2\frac{e^{2i(\xi-\zeta)}}{e^{2l\alpha\cos\zeta}+Re^{-2i\zeta}}\Big\rbrace dP.
\end{equation}

Note that the compact expression (\ref{dF}) is equivalent to Eqs. (\ref{eq:dfz}) and (\ref{eq:dfy}).

\section*{Appendix B}

\begin{figure}
\includegraphics[width=\linewidth]{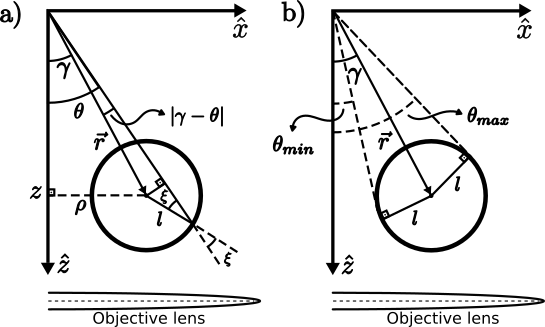}
\caption{a) Spherical particle placed out from the focus, along with the definition of the coordinate system. See text for details. Fig. b) illustrates the fact that only those rays between $\theta_{min}=\mathrm{Max}[0,\gamma-\arcsin(l/r)]$ and $\theta_{max}=\gamma+\arcsin(l/r)$ reach the sphere.}
\label{figure33}
\end{figure}

Consider an optical tweezers made of a highly focused laser beam, whose intensity is modulated as a Gaussian profile. Let the  $z$-axis be placed along the optical axis, pointing toward the objective lens (Fig. \ref{figure33}-a)). If we choose the origin of the coordinate system to be at the laser focus, it is well-known that power increment $dP$ of a single light ray can be written as \cite{Rocha}:%
\begin{equation}
dP=\frac{2P_t}{\pi \sigma^2}\exp\left(\dfrac{-2f^2\sin^2\theta}{\sigma^2}\right)f^2\sin\theta\cos\theta d\theta d\phi,
\end{equation}%
where $\theta$ is the angle between the light ray and the optical axis of the Gaussian beam, and $\phi$ is the usual azimuthal angle. The limits of integration for $\theta$ and $\phi$ will be discussed in details below. $P_t$ is the total power before the laser enters the objective, $\sigma$ is the beam waist, and $f$ is the focal length.

Fig. \ref{figure33}-a) depicts the spherical particle (radius $l$) at a distance $z$ below the focal plane, at position $\vec{r}$. The vector $\vec{d}$ represents the position where the incoming light ray reaches its surface. The $x$-$z$ plane was chosen to cross the center of the sphere, and $\gamma$ is the angle between $\vec{r}$ and the optical axis. Let $\rho$ be the radial distance from the optical axis to the geometric center of the sphere.

Writing $\hat{y'}$, $\hat{z'}$, and all the variables as functions of $\theta$ and $\phi$, and integrating Eq. (\ref{dF}), one obtains the total radiative force exerted on the bead by all light rays reaching its surface. The vectors $\vec{r}$ and $\vec{d}$ can be written as \cite{Rocha}:%
\begin{equation}
\begin{aligned}
\vec{r}\,=\, &r(\sin\gamma,0,\cos\gamma),\\
\vec{d}\,=\,&d(\sin\theta\cos\phi,\sin\theta\sin\phi,\cos\theta),
\end{aligned}
\end{equation}%
where%
\begin{equation}
\begin{split}
d=\sqrt{l^2-r^2+r^2(\sin\gamma\sin\theta\cos\phi+\cos\gamma\cos\theta)^2}\,+\\
+\,r(\sin\gamma\sin\theta\cos\phi+\cos\gamma\cos\theta).
\end{split}
\end{equation}%
The $\xi$ angle can be written as%
\begin{equation}
\xi=\arccos\left(\dfrac{d^2+l^2-r^2}{2ld}\right).
\end{equation}%
The unitary vectors $\hat{z'}$ and $\hat{y'}$ are given by:

\begin{equation}
\begin{aligned}
\hat{z'}\,=\,&(-\sin\theta\cos\phi,-\sin\theta\sin\phi,-\cos\theta),\\
\hat{y'}\,=\,&\frac{\hat{z'}\times (\vec{r}\times \hat{z'})}{\vert \hat{z'}\times (\vec{r}\times \hat{z'})\vert}.
\end{aligned}
\end{equation}

\begin{figure}
\includegraphics[width=.65\linewidth]{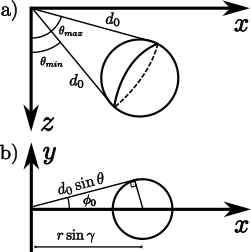}
\caption{Illustration of the procedure used to calculate the limits of integration for $\phi$. In a) one sees the curve on the surface of the sphere corresponding to the tangent angles of incidence. The rays tangency the sphere at a fixed distance $d_0$ from the origin for all values of $\phi_0(\theta)$. b) shows a cross-section parallel to the focal plane, with the coordinates of a tangent ray, together with the limit of integration for the azimuthal angle, $\phi_0$.}
\label{img:limitphi}
\end{figure}

Note that only a portion of the light rays leaving the objective lens actually reaches the sphere, whenever it is placed out of the focus of the laser beam. The limits of integration for $\theta$ are given by (see Fig. \ref{figure33}-b)):

\begin{equation}
\begin{aligned}
\theta_{min}\,=&\,\textrm{Max}[0,\gamma-\arcsin(l/r)],\\
\theta_{max}\,=&\,\textrm{Min}[\theta_0,\gamma+\arcsin(l/r)],
\end{aligned}
\end{equation}%
where $\theta_0$ is the critical angle for the glass-medium interface at the coverslip ($\theta_0=1.109$ rad for glass-water interface).

To determine the limits of integration for $\phi$, notice that they correspond to the situations where the light rays are tangent to the surface of the bead (Fig. \ref{img:limitphi}-a)). This defines a curve at the surface, whose equation can be obtained by subjecting the equation of the sphere to the constraint $d=d_0\equiv d(\theta=\gamma+\arcsin(l/r),\phi=0)$. Therefore, this curve must satisfy $(x-x_0)^2+(y-y_0)^2+(z-z_0)^2=l^2$, where (see Fig. \ref{img:limitphi}-b)):%

\begin{equation}\left\{
\begin{array}{lll}
x=d_0\sin\theta\cos\phi,& & x_0=r\sin\gamma,\\
y=d_0\sin\theta\sin\phi,& & y_0=0,\\
z=d_0\cos\theta,& & z_0=r\cos\gamma.
\end{array}
\right.
\end{equation}

As a result, $\phi_{min}(\theta)=-\phi_0(\theta)$ and $\phi_{max}(\theta)=\phi_0(\theta)$, where
\begin{equation}
\phi_0(\theta)=\arccos\left[\left(\frac{d_0^2+r^2-l^2}{2rd_0}-\cos\theta\cos\gamma\right)\csc\theta\csc\gamma\right].
\end{equation}

\begin{figure}[h]
\includegraphics[width=.75\linewidth]{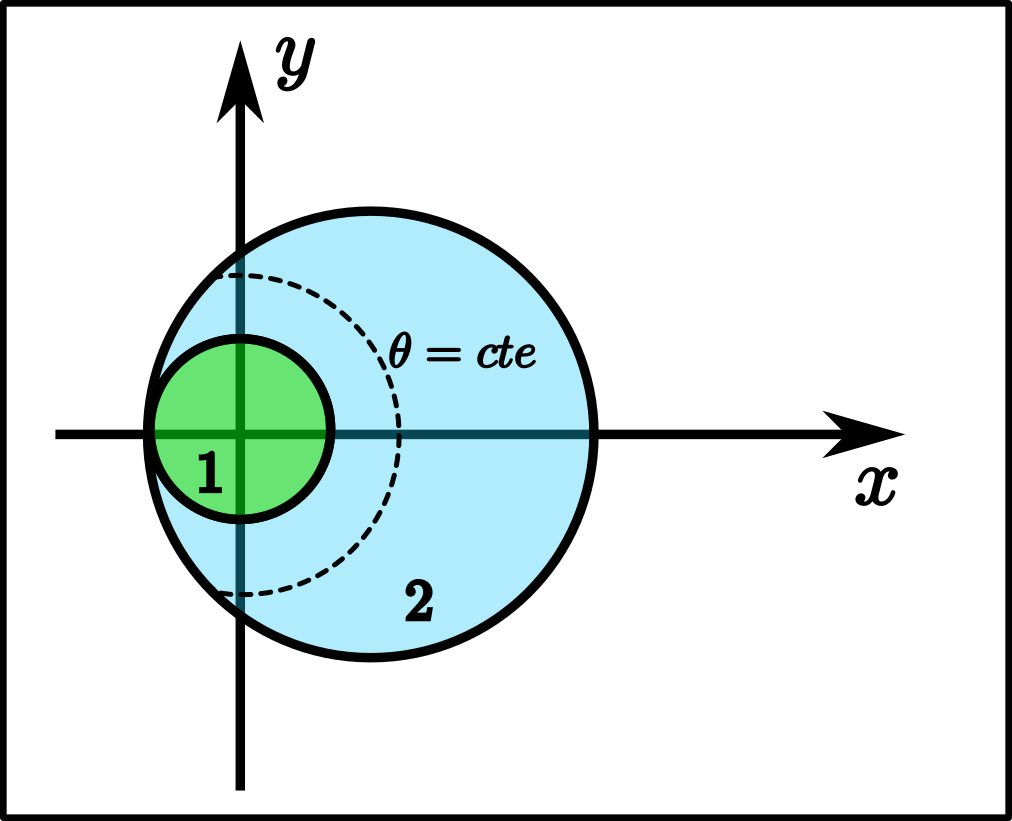}
\caption{(Color online) Cross-section parallel to the focal plane for the situation where $\rho<l$. If $\theta<\vert \gamma-\arcsin (l/r)\vert$ (region 1-green), the light ray crosses the sphere for all values of $\phi$. However, if $\theta>\vert \gamma-\arcsin (l/r)\vert$ (region 2-blue), only those rays parametrized by $\phi \in [-\phi_0,\phi_0]$ cross the sphere.}
\label{im:regions}
\end{figure}

This is valid when $\rho>l$, and the $z$-axis does not intercept the sphere. However, if $\rho<l$, one needs to be more careful with the limits of integration. Fig. \ref{im:regions} shows a cross section parallel to the focal plane in such case. For $0<\theta < \vert \gamma-\arcsin(l/r) \vert$ (region 1), we have $-\pi \leq \phi \leq \pi$, while for $\theta > \vert \gamma-\arcsin(l/r)\vert$ (region 2) we have $-\phi_0(\theta)<\phi<\phi_0(\theta)$. Therefore, in this case
\begin{equation}
\phi_{max}=\left\{%
\begin{array}{ll}
\pi\,,\quad &0<\theta < \vert \gamma-\arcsin(l/r) \vert\\
\phi_0(\theta)\,,\quad &\vert \gamma-\arcsin(l/r) \vert<\theta<\theta_{max} .
\end{array}\right.
\end{equation}%
It is easy to see that $\phi_{min}=-\phi_{max}$.

\bibliography{AbsorptionOT_bibtex}


\end{document}